\title{Decadal_White_Paper}

\documentclass[letterpaper,12pt]{article}

\usepackage{pdfpages}
\usepackage[super]{natbib} 
\usepackage{apjfonts} 
\usepackage{overcite}
\usepackage{graphicx} 
\usepackage{amssymb,amsmath} 
\usepackage{color} 
\usepackage{enumitem} 
\usepackage{wrapfig} 
\usepackage{pdflscape} 
\usepackage{afterpage} 
\usepackage{hyperref}

\newcommand{\xray}{\mbox{X-ray}}
\newcommand{\ictxs}{\mbox{IceCube-170922A}}
\newcommand{\txsqso}{\mbox{TXS~0506+056}}
\newcommand{\cgs}{\mbox{erg~cm$^{-2}$~s$^{-1}$}}

\usepackage{color}
\usepackage{graphics,graphicx}
\usepackage{soul}
\usepackage{times}
\usepackage{subfig}
\usepackage{setspace}
\usepackage[font={footnotesize}]{caption}
\usepackage[font=singlespacing]{caption}
\usepackage{multicol}
\usepackage{apjfonts}
\usepackage{caption}

\usepackage{etoolbox}

\makeatletter
\patchcmd{\@footnotetext}{\footnotesize}{\scriptsize}{}{}
\makeatother

\makeatletter
\renewcommand{\section}{\@startsection%
{section}{1}{0mm}{-\baselineskip}%
{0.5\baselineskip}{\normalfont\normalfont\bfseries}}%
\makeatother

\begin{document}
\pagestyle{plain}
\pagenumbering{arabic}

\includepdf[pages=-]{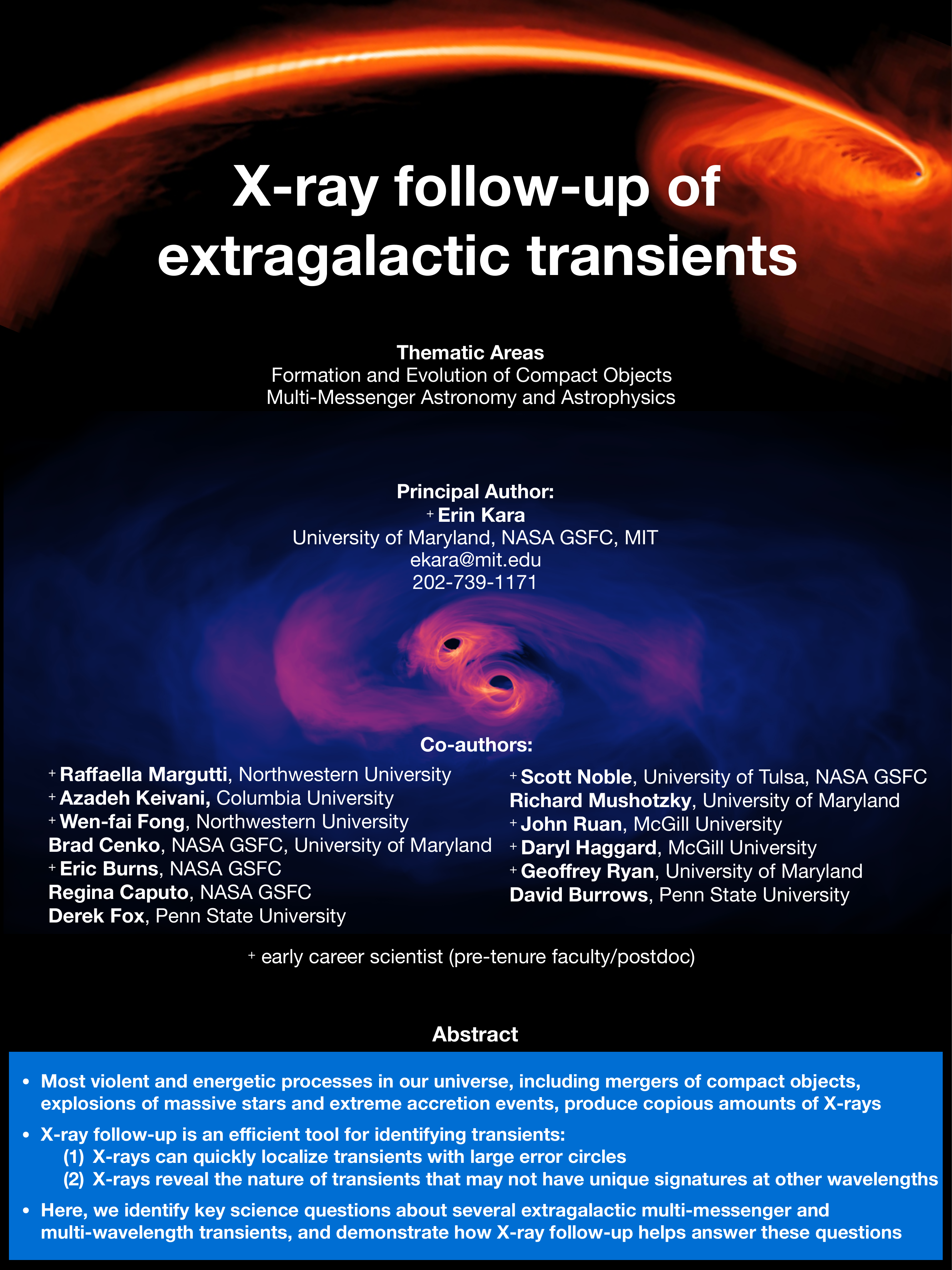}

\begin{center} 
\bfseries\uppercase{X-ray follow-up of extragalactic transients}
\end{center}
\vspace{-0.3cm}

\vspace{-0.4cm}
\section{Introduction}
\vspace{-0.1cm}
Over the last 50 years X-ray observations have played a major role in Time Domain Astronomy: from the discovery of new galactic black holes, giant stellar flares, tidal disruption events, and flares from SgrA*, to the existence of super-Eddington accretors, the discovery of magnetars, and millisecond X-ray pulsars.  Almost all these phenomena were true `discoveries', in that they were not predicted. Even now, after 50 years of study, entirely new classes of  transient sources are still being discovered in almost every wavelength band. The future of Time Domain Astronomy is only limited by our imagination.

The ASTRO2010 Decadal Survey ranked the LSST as the highest priority project for ground-based astronomy, indicating that time-domain astronomy is an essential component in understanding our universe. The next decade will see vast increases in the rate of transient discovery. Expectations are that LSST will see thousands to tens of thousands of transients per night. In the radio, SKA will find tens of transients per night. In X-rays, the {\em eROSITA} all-sky survey (2019 launch) will provide a 30-fold improvement in sensitivity, discovering faint transients, and producing a baseline against which we will compare future observations. Upgrades to existing ground-based GW facilities will enable detections of tens to hundreds of neutron star-black hole and binary neutron star mergers per year. Finally, by the 2030s, {\em LISA} will measure GW signals from a few to a hundred supermassive black hole mergers annually\citep{klein16}.  

Large time domain surveys require multi-wavelength follow-up to understand the nature of transients and their physical properties. Here we focus on the necessity for X-ray observations in answering several key astrophysical questions.

\vspace{-0.2cm}
\section{Tidal Disruption Events}
\vspace{-0.2cm}
{\it \textbf{What does super-Eddington accretion onto massive black holes look like?}}

Super-Eddington accretion is one of the proposed solutions to how the first supermassive black holes grew\citep{volonteri05,madau14}, but our understanding of it is not yet on firm footing. Although the overall radiative efficiency and luminosity are still debated\citep{sadowski16,jiang14}, all simulations show a geometrically and optically thick accretion flow with strong, fast outflows. Amongst the best examples of super-Eddington accretion are tidal disruption events (TDEs) in the local universe.

Roughly one star per galaxy every 10 thousand years\citep{stone16} is disrupted by strong tidal forces from the central supermassive black hole\citep{rees88,phinney89}. The fallback rate of the stellar debris is initially highly super-Eddington for $M_{\mathrm{BH}}<10^{7} M_{\odot}$ and drops with time. If the stellar debris can lose its orbital energy quickly (i.e. through stream-stream collisions\citep{shiokawa15,piran15,dai15}), then the accretion rate onto the black hole can also exceed the Eddington limit\citep{lodato11}. TDEs are particularly interesting test cases because we can directly follow the transitions from super-Eddington to sub-Eddington accretion in an individual system on timescales of months. Understanding the X-ray component of TDEs is key, as this emission comes from the innermost regions, where massive outflows and relativistic jets can be launched. 

Recently, evidence for $v=0.2-0.5c$ ultrafast outflows has been seen in 3 TDEs\citep{lin15,kara16,kara18} through observations of blueshifted spectral lines in the X-ray. Moreover, in one source, late-time follow-up revealed that the outflow `shuts off' as the luminosity drops, as predicted from standard accretion theory\citep{kara18}. Unfortunately, we currently lack fast-slewing, high-effective area X-ray instruments that can efficiently track the detailed spectral evolution of these unique accretion flows. 

{\it \textbf{X-ray Identification}}: LSST is expected to find up to 1,000 TDEs per year\citep{vanvelzen11} (compared to the current rate of 1-2 TDEs/year). With this influx of optical events, an ongoing question will be: is this nuclear transient a TDE flare or simply due to AGN activity? X-ray follow-up will be key because X-rays are one of the cleanest indicators of TDE activity. AGN show hard X-ray emission from the X-ray corona, but nearly all of the X-ray emitting (non-jetted\citep{burrows11}) TDEs lack this hard X-ray component, and instead show a soft $10^{5-6}$~K thermal spectrum\citep{auchettl17}. Not all optically-selected TDEs emit in X-rays, and there is much debate on whether this suggests that there are several types of TDEs (differentiated, in part, by how efficiently orbital energy of the stellar debris can dissipate) or if the observed TDEs can be unified in one picture, driven largely by orientation affects\citep{dai18,curd19}. To observe a TDE with $L_{\mathrm{X}}=10^{42}$~erg~s$^{-1}$ at z=2 requires a sensitivity of $1\times 10^{-16}$~erg~cm$^{-2}$~s$^{-1}$, which is obtainable with modest exposures with {\em Athena}, {\em Lynx} or {\em AXIS}. Equally important to X-ray follow-up of optically-discovered TDEs will be optical/UV follow-up of X-ray discovered TDEs, as found with {\em eROSITA} or the Wide Field Monitor on {\em STROBE-X}. Understanding the connection between UV, optical and X-ray emission is key to resolving the basic structure of TDEs. 

\begin{figure}[t]
  \begin{minipage}[c]{0.67\textwidth}
    \includegraphics[width=\textwidth]{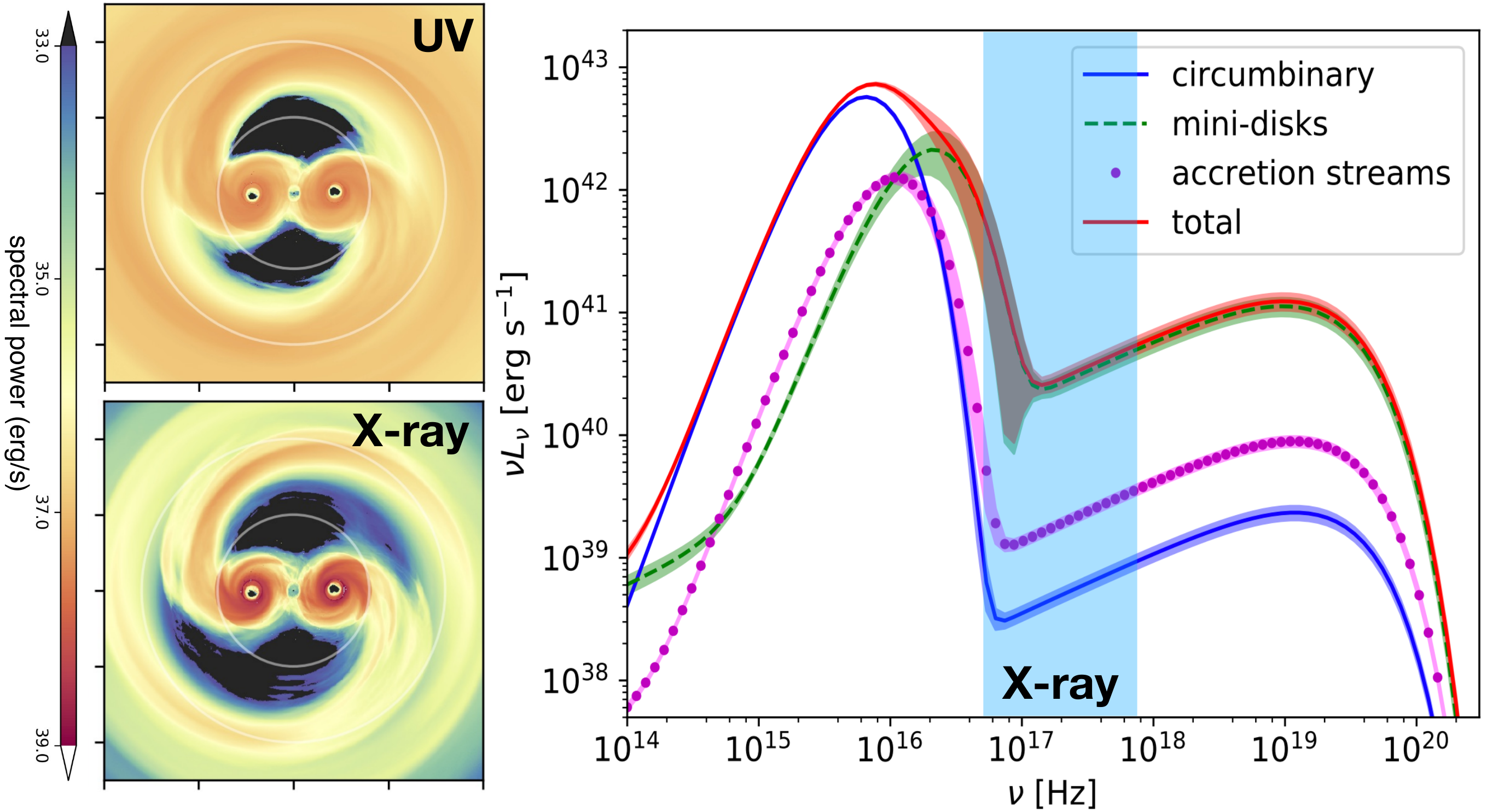}
  \end{minipage}\hfill
  \begin{minipage}[c]{0.3\textwidth}
    \caption{\small{
       {\em Left:} Simulated images of supermassive black hole binaries in UV and X-ray bands, demonstrating that the X-ray emission is dominated by minidisc emission, while longer wavelengths are a combination of minidiscs, gas streams and the circumbinary disc. {\em Right:} The corresponding SED. Adapted from d'Ascoli et al., 2018.
    }} \label{fig:SMBHB}
  \end{minipage}
  \vspace{-0.4cm}
\end{figure}

\vspace{-0.2cm}
\section{{\em LISA} events: Supermassive black hole mergers}
\vspace{-0.2cm}
{\it \textbf{ How do supermassive black holes merge? }}

A fundamental prediction of hierarchical structure formation models is that supermassive black holes will grow by successive coalescences following the merger of their host halos. After the two galaxies merge, a supermassive black hole binary will form by stellar and gas dynamical processes, and then will coalesce due to the emission of low frequency gravitational waves\citep{mayer07} (GWs). {\em LISA}, a space-based GW detector scheduled to launch in the early 2030s, will detect black hole inspirals and mergers from $10^{3-7} M_{\odot}$ black holes up to a redshift of z=20 \citep{amaro-seoane17}. A typical {\em LISA} event will be $10^{6} M_{\odot}$ black hole mergers at the peak of star formation at z=2. Supermassive black hole binaries (SMBHBs) are promising multi-messenger events, as accretion discs will likely form during the interaction of gas rich galaxies\citep{armitage02}.

The gas dynamics around SMBHBs and the subsequent electromagnetic signatures are a subject of active research. It is generally agreed that SMBHB systems with large mass ratios exist in a circumbinary disc\citep{pringle91,macfadyen08,noble12,shi12} 
which feeds mini-discs around the two black hole\citep{bowen18,ryan17}. 
These mini-discs are thought to produce thermal emission in soft X-rays and a hard X-ray continuum at luminosities close to the Eddington limit\citep{dascoli18,roedig14}.  
While the broadband SED and variability timescales are still highly uncertain, it is likely that the X-rays will be dominated by the mini-disc's emission (with significantly less from the circumbinary disc or gas streams). Therefore X-rays will likely be the cleanest tracer of the black hole orbital frequency (i.e. half the gravitational wave frequency). Moreover, because X-rays cleanly trace mini-disc emission, Doppler beaming effects\citep{dorazio15} will likely be strongest in this band. 

{\it \textbf{X-ray Identification}}: Whatever the exact SED for a SMBHB, it is likely that these candidates will be AGN-like.  Even by conservative estimates, where these black holes radiate at the 10\% Eddington luminosity, and X-rays only account for 10\% of the bolometric luminosity (as in typical AGN), then at a redshift of z=2 for a typical $10^{6} M_{\odot}$ merger, we expect a flux of $3\times10^{-16}$~erg~cm$^{-2}$~s$^{-1}$. If the source is moderately obscured ($N_H=5 \times 10^{22}$~cm$^{-2}$, as is seen for many dual AGN\citep{koss18}), the estimated flux at z=2 will be $1\times 10^{-16}$~erg~cm$^{-2}$~s$^{-1}$, still well within the grasp of some proposed X-ray missions, such as {\em AXIS}, {\em Lynx} and {\em Athena}. More optimistic estimates of similar mass SMBHB inspirals with higher radiative efficiency (L=$L_{\mathrm{Edd}}$) at z=1 could be detected by {\em TAP}. {\em LISA} will have 1 GW event per year with an error circle of $<1$~deg$^2$, with a distance measure within 10\%, and $\sim$5 events per year with error circles of 10~deg$^2$ or more\citep{lang08}. The ``warning'' time, when a position can be reasonably estimated, is $\sim$16-35 days\citep{haiman17}. High-sensitivity X-ray missions can tile the 1~deg$^2$ error circle in a reasonable number of pointings, in order to provide the community with a handful of high probability GW counterpart candidates\citep{delcanton19}. The SMBHB can then be identified by orbital periodicities or perhaps by an unusual and quickly evolving broadband SED. 

\vspace{-0.2cm}
\section{Blazars as Neutrino Counterparts}
\vspace{-0.2cm}
{\it \textbf{ What is the particle composition of relativistic jets?}}

In 2013 the IceCube Neutrino Observatory announced its discovery of a near-isotropic flux of high-energy cosmic neutrinos \citep{2013PhRvL.111b1103A,2013Sci...342E...1I}. 
While the average flux, spectrum and distribution of neutrino flavors has been determined, their astrophysical origins remain uncertain. Searches for point sources, neutrino ``hot spots,'' an excess of neutrinos along the Galactic plane or toward the Galactic center, or excesses of neutrinos in association with any of a range of pre-selected source candidates, continue to yield null results. 

The sole exception to date was provided via rapid-response multi-messenger follow-up observations of the ``\ictxs'' high-energy neutrino alert.
The worldwide campaign of follow-up observations for this alert provided many multiwavelength detections and evidence of extraordinary {\em Fermi}-detected gamma-ray activity by the BL~Lac-type blazar \txsqso, near the center of the neutrino localization.
This blazar was ultimately shown to be the source of the \ictxs\ neutrino with $>$3-sigma confidence \citep{2018Sci...361.1378I}.
Further analysis suggests that the source generated high-energy neutrinos over a 6-month flare, which can explain thirteen IceCube events over backgrounds. 

Understanding the X-ray behavior of neutrino counterparts is vital for modeling the particle content in the jet because X-rays are produced through electromagnetic cascades from hadronic interactions that produce neutrinos. The short scattering lengths for $e^{+}e^{-}$ and gamma-rays mean that the neutrinos maintain their high energy, while much of the EM emission emerges in the X-ray band. This means that upper limits on the neutrino fluence are set by the X-ray fluence. For the case of TXS~0506+056, the {\em Swift}-XRT observations tightly constrained the maximum predicted neutrino flux for a leptonic model explained in~\citep{2018ApJ...864...84K}.
\xray\ data offers the most constraining limit on the photon emission type (e.g.\ synchrotron self-Compton vs. external inverse-Compton) at high energies. 

{\it \textbf{X-ray Identification}}:
The successful campaign of \ictxs\ indicates that rapid electromagnetic follow-up of neutrino alerts may be the best way to identify and characterize neutrino counterparts. The current IceCube configuration provides $\sim 10$ neutrino alerts per year, and constrain a 50\% error circle of $1-2 \deg^2$. The next generation IceCube detector, planned for the late 2020s, will provide a tenfold increase in sensitivity and 5x better spatial resolution. Pointed X-ray telescopes with fast response times of a few hours can follow-up neutrino alerts and look for flaring activity, compared to baselines provided by {\em eROSITA}, which will scan the entire X-ray sky down to a sensitivity of $10^{-14}$~\cgs.

\begin{figure}[t]
  \begin{minipage}[c]{0.67\textwidth}
    \includegraphics[width=\textwidth]{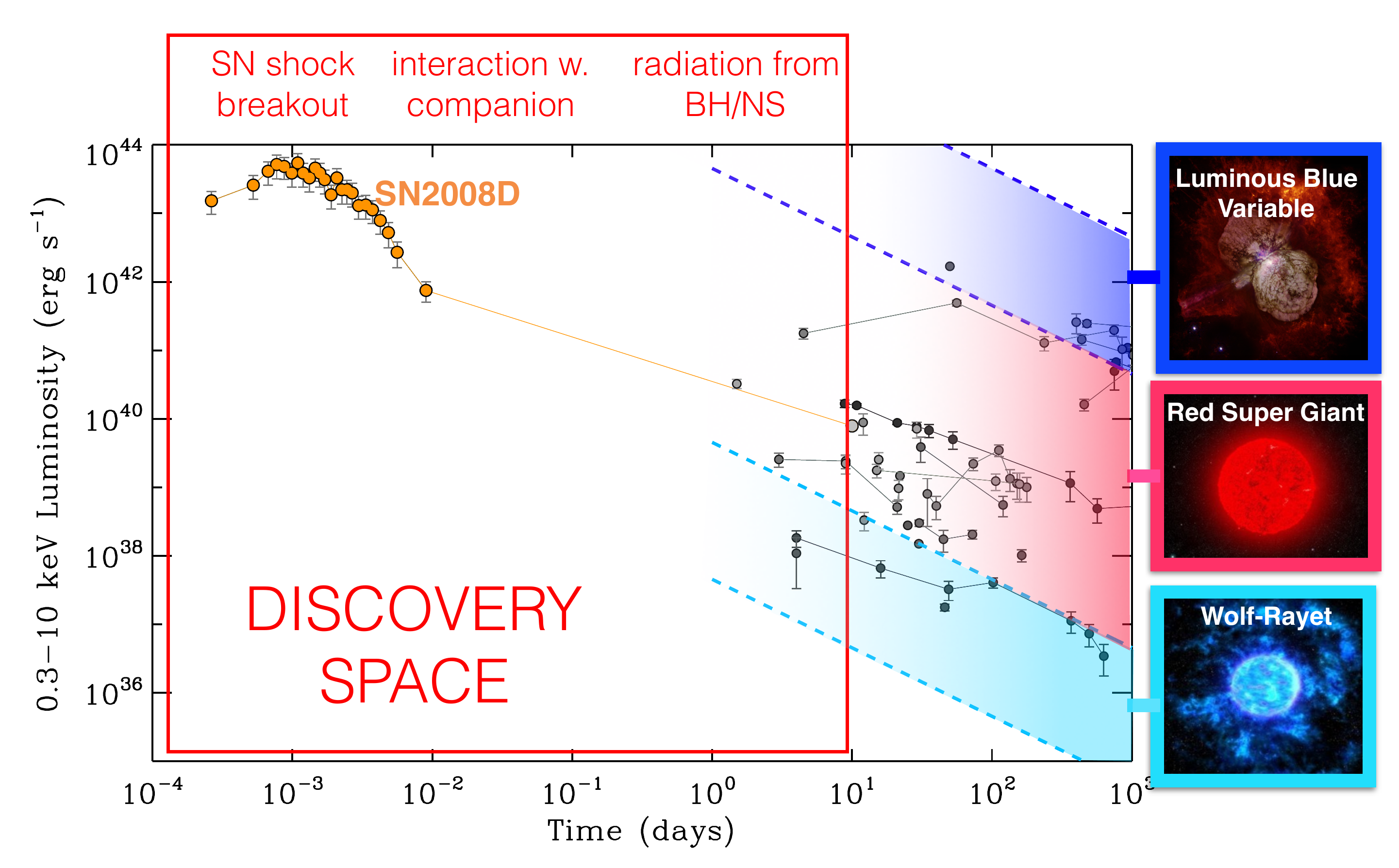}
  \end{minipage}\hfill
  \begin{minipage}[c]{0.3\textwidth}
    \caption{\small{
       The detected X-ray emission from SNe to date. There is pristine territory of exploration, at $t<2$ weeks from stellar explosion, which will unveil the currently elusive progenitors of core-collapse hydrogen-poor SNe and Type-Ia SNe (which together comprise $\sim50$\% of known stellar deaths). Later observations ($t>2$ weeks) of X-rays from SN shock interaction with the medium will connect stellar deaths to the different stellar progenitors.
    }} \label{fig:plot}
  \end{minipage}
  \vspace{-0.6cm}
\end{figure}

\vspace{-0.2cm}
\section{Supernovae}
\vspace{-0.2cm}
{\it \textbf{ What are the progenitors of H-poor supernova and how do massive stars evolve to their deaths?\\}}
The structure of stars at the time of explosion and their recent mass-loss history in the last thousands of years before stellar death are two of the least understood aspects of stellar evolution\citep{Smith14,Waxman16}.
Equally embarrassing is our complete lack of knowledge of which stars are progenitors of $>50\%$ of supernovae (SNe), including super-luminous SNe\citep{Quimby11}, Gamma-Ray Burst SNe, which produce the most relativistic jets known \citep{Hjorth12}, as well as SNe of Type Ia, which have been employed as cosmic ladders to reveal the accelerating Universe \citep{Riess98}. This lack of understanding also impacts estimates of the initial stellar mass function in galaxies and star formation through cosmic time\citep{Smith14}, in addition to severely limiting predictive power of current theories of stellar evolution and the cosmological use of Type Ia SNe \citep{Maoz14}. 

X-ray observations of young SNe can fill this knowledge gap in the following four ways:
\vspace{-0.2cm}
\begin{enumerate}[leftmargin=*]
\setlength{\itemsep}{-2pt}
\vspace{-0.1cm}
\item SN shock breakout from the surface of a massive envelope-stripped star may be the best indicator of the stellar progenitors of hydrogen-poor SNe\citep{Waxman16}. 
This phenomenon produces a flash of X-rays with temperature $\sim0.1-10$ keV, duration from tens of seconds to minutes and energy $E\sim10^{43}-10^{45}$ erg. There has only been one (serendipitous) detection thus far\citep{Soderberg08} (see SN2008D in Fig. \ref{fig:plot}).

\item SN shock interactions with the companion star a few hours after explosion 
provide direct insight into the nature of the companion (WD or non-degenerate star) in Type-Ia SNe\citep{Kasen10}. This new and totally unexplored X-ray technique could significantly reduce the systematic uncertainties in distances that afflict current dark energy constraints.

\item SN shock interactions with the circumstellar medium\citep{Fransson96} produce copious amounts of X-ray emission for several weeks after the explosion\citep{Margutti14,Margutti17}. The circumstellar environment imprints information about the progenitor mass loss history centuries to thousands of years before the collapse \citep{Dwarkadas12}.

\item Massive stellar explosions are the known factories of compact objects in the Universe. Yet, which SN produces which compact object (NS or BH?) remains an open question. With the Astronomical Transient AT2018cow, early-time ($t<2$ weeks) coordinated soft and hard X-ray observations provided a new window of exploration into the very first moments after the formation of a BH or NS \citep{Margutti19}.
\end{enumerate}
\vspace{-0.1cm}
\noindent {\it \textbf{X-ray Identification:}} The extremely short-lived nature of the shock break out and companion interaction emission ($\Delta t<$ hrs) have so far prevented a systematic exploration of the X-ray phase space of cosmic stellar explosions (Fig. \ref{fig:plot}).  SN shock breakout and shock interactions with the companion star (detected with sensitive wide-field X-ray instruments, like the {\em TAP} with a $1\rm{deg^{2}}$ FoV) will be associated with SNe detected later by wide field, deep optical transient surveys like ZTF and LSST. Optical surveys will also, of course, discover new SNe, which can be be followed up in X-rays on timescales of hours to weeks to measure the circumstellar medium interaction signal, which is mostly below the sensitivity limit of currently existing  ``time-domain machines'' like {\em Swift}. A vast discovery phase space is in principle available to sensitive, fast follow-up X-ray instruments. 

\vspace{-0.2cm}
\section{LIGO/VIRGO events: Neutron Star Mergers and Short $\gamma$-ray Burst Afterglows}
\vspace{-0.2cm}
{\it \textbf{What is the nature of relativistic outflows from NS mergers, and what is the remnant compact object?}}\\
Short gamma-ray bursts (SGRBs) are 
amongst the most luminous EM transients in the universe.  Their progenitors are compact object mergers, involving two neutron stars or a neutron star and black hole (NS-NS/NS-BH). Such systems are powerful sources of GW emission, and are sites of heavy element production\citep{cbk+17,kmb+17,met17,pdb+17}. In the 2020s, as LIGO/Virgo \citep{LIGO,Virgo} reach design sensitivity and future facilities (KAGRA \citep{KAGRA}, aLIGO Plus\citep{Aplus}) come online, we will enter a golden era for studies of compact objects and their mergers.

 Deep, high-cadence X-ray observations play a critical role in delineating the nature of outflows in neutron star mergers. One of the key revelations to emerge from 
 the first NS-NS merger, GW170817\citep{gw170817}, was evidence for a relativistic jet launched from the merger\citep{amb+18,Mooley18,WuMacFadyen18,Nynka18,Ghirlanda19}. In particular, the initial {\em Chandra} detection of the brightening X-ray emission provided the first indication that its non-thermal afterglow was unlike any other short GRB. Continued X-ray monitoring over ~500 days helped distinguish between competing models for the afterglow, which is now thought to be a structured jet that is viewed off-axis; this effort combined used $\sim 850$~ksec of {\em Chandra} time (past and planned). 

 Looking forward, X-ray observations of a sample of NS-NS mergers will be critical for gravitational wave cosmology, by breaking the degeneracy between the distance and the viewing angle of the binary, and for the physics of the central engine, by constraining the energetics and collimation of the outflow. Furthermore, X-ray observations of future gravitational wave events will provide insights to new phenomena, such as the anticipated discovery of a NS-BH merger, and X-ray emission from a stable magnetar remnant produced in a NS-NS merger.

{\it \textbf{X-ray Identification:}} 
By the early 2020s, most GW events will be found at a distance of $\sim~200$~Mpc. If these events are like GW170817, their X-ray afterglow will have $F_{\mathrm{x}}\sim 8\times 10^{-16}$ erg~cm$^{-2}$~s$^{-1}$, below the sensitivity of Chandra for a deep 100 ks exposure. Moreover, GW170817 was at a projected offset of $\sim 2$~kpc from an AGN at its host center, translating to an offset of $\approx 0.5''$ for a more distant merger at 200~Mpc. We expect NS-NS mergers to occur at a range of offsets from their host galaxies \citep{fb13,bpb+06}, so sub-arcsecond angular resolution is key to avoiding source confusion for low-offset events. In the next decade, it is critical to have sensitive, rapid, and high angular resolution X-ray observations of GW sources to understand whether some, most or all NS-NS and NS-BH mergers produce similar relativistic emission.

\vspace{-0.4cm}
\section{Serendipitous Science}
\vspace{-0.2cm}
{\it \textbf{What surprises await in the transient soft X-ray sky?}}\\
While X-ray follow-up of transients discovered at other wavelengths will be critical in the coming decade, recent results have also demonstrated that the transient (soft) X-ray sky remains a rich \textbf{discovery} 
space.  In addition to the serendipitous detection of shock breakout emission from SN2008D
\citep{Soderberg08}, more recently new and as-yet-unexplained X-ray transients have been discovered in both nearby \citep{Jonker+2013,Irwin+2016} and distant \citep{Bauer+2017} galaxies.  Theoretical models also predict a rich phenomenology of (as-yet-unobserved) X-ray transients, from ``orphan'' (i.e., off-axis) afterglows of gamma-ray bursts to high-energy counterparts to the new population of fast radio bursts.  X-ray survey facilities that optimize observational cadences to sample a wide range of time scales will surely uncover new and unexpected transient classes in the coming decade.

\vspace{-0.4cm}
\section{Conclusion and future outlook}
\vspace{-0.2cm}
In the past decade, X-ray follow-up of extragalactic transients has largely been driven by {\em Swift}-XRT because of its fast slew and settling time, and because of its willingness to take short snapshots and follow-up quickly and repeatedly on exciting multi-wavelength and multi-messenger triggers. {\em XMM-Newton} has also played a vital role in understanding these phenomena thanks to its large effective area (10x {\em Swift-XRT}), which enables detailed spectroscopic and timing analyses. Finally, the high spatial resolution of {\em Chandra-ACIS} is invaluable for identifying sources in crowded fields. 

With the advancements of time domain surveys, gravitational wave facilities and neutrino detectors in the next decade, corresponding follow-up capabilities with \textbf{fast slew, high effective area and high spatial resolution} are necessary. Thanks to breakthroughs in X-ray mirror and detector technologies, these three time domain specifications are possible in moderately priced missions launching the late 2020s.  

\vspace{-0.3cm}
\begin{figure}[h]
\includegraphics[width=\textwidth]{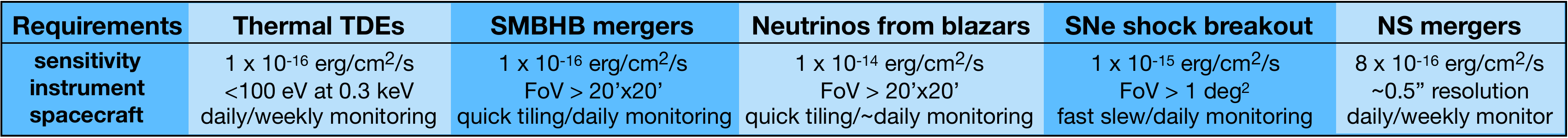}
\vspace{-0.8cm}
\caption*{\footnotesize{The requirements for a future 0.3--10~keV X-ray telescope that would enable the science discussed in this white paper. }}
\end{figure}

\newpage

\section*{Figure Credit}
Title Page: Tidal disruption event simulation by James Guillochon. Binary supermassive black hole simulation by Scott Noble and visualized by NASA GSFC Press office. Fig~\ref{fig:SMBHB} adapted from d'Ascoli et al., 2018. Fig~\ref{fig:plot} by Raffaella Margutti.

\bibliographystyle{aasjournal}
\bibliography{wp}

\end{document}